\documentclass[twocolumn,showpacs,preprintnumbers,amsmath,amssymb]{revtex4}

\usepackage{graphicx}
\usepackage{dcolumn}
\usepackage{bm}

\newcommand{\gsim}{\raisebox{0.2ex}{$\ > \kern -1.05em%
        \raisebox{-1.1ex}{$\sim$}\ $}}  

\begin{document}

\preprint{KANAZAWA/02-09}

\title{Nonperturbative renormalization-group approach for
quantum dissipative systems}

\author{Ken-Ichi Aoki}
 \email{aoki@hep.s.kanazawa-u.ac.jp}
\author{Atsushi Horikoshi}%
 \email{horikosi@hep.s.kanazawa-u.ac.jp}
 \altaffiliation[Present address: ]
{Japan Science and Technology Corporation, 
and Department of Chemistry, Faculty of Science, 
Nara Women's University,
Nara 630-8506, Japan; 
e-mail: horikosi@cc.nara-wu.ac.jp
} 
\affiliation{%
Institute for Theoretical Physics, Kanazawa University, \\
Kakuma-machi Kanazawa 920-1192, Japan
}%

\date{\today}

\begin{abstract}
We analyze the dissipative quantum tunneling 
in the Caldeira-Leggett model 
by the nonperturbative renormalization-group method.
We classify the dissipation effects by 
introducing the notion of effective cutoffs.
We calculate the localization susceptibility 
to evaluate the critical dissipation for the 
quantum-classical transition. 
Our results are consistent with previous semiclassical arguments, 
but give considerably larger critical dissipation.
\end{abstract}

\maketitle

\section{Introduction}
There has long been great interest 
in classical or quantum systems of a few degrees of freedom 
coupled with an external environment,
both theoretical and experimental. 
It has something to do with very fundamental questions 
about quantum decoherence, nonequilibrium open systems, etc.
Recently, modern experiments using mesoscopic systems
have opened the possibility of directly measuring and investigating 
these fundamental issues.
Also, it will provide key information 
for nanophysics and nanotechnology in the near future. 
It is difficult to consider the whole influence of the environment, 
and here we concentrate on the dissipative behaviors, since 
they are typical characteristics of open systems.
\par
A method to derive the dissipative behavior 
from microscopic nondissipative theory
was proposed by Caldeira and Leggett \cite{cl1,cl2}. 
Their model consists of two sorts of degrees of freedom, 
the target system and the environment.
The action is expressed as follows:
 \begin{eqnarray}
{S}[~q,\{x_{\alpha}\}]
&=&
\int d{t}~\left\{
\frac{1}{2}M{\dot q}^{2} - V_0(q)
\right.
\nonumber\\
+\sum_{\alpha}&&\!\!\!\!\!\!\!\!\!
\left.\left[~\frac{1}{2}m_{\alpha}{\dot x_{\alpha}}^{2}
-\frac{1}{2}m_{\alpha}\omega_{\alpha}^{2}x_{\alpha}^{2}
-q C_{\alpha}x_{\alpha}\right]\right\}, \label{(1)}
\end{eqnarray}
where $q(t)$ is the variable of the target system 
in a potential $V_{0}(q)$, and 
$x_{\alpha}(t)$ are the harmonic oscillators representing 
the environmental degrees of freedom.
The target system variable is coupled linearly 
to each oscillator with strength $C_{\alpha}$.
If we set the parameters 
$m_{\alpha}$, $\omega_{\alpha}(>0)$, and $C_{\alpha}$ 
in an appropriate way, 
and eliminate the environmental variables 
with the proper boundary condition,
a dissipative term 
(for example, the Ohmic dissipation $\dot{q}$) 
arises in the {\it classical} 
effective equation of motion of $q(t)$.
\par
In this article, we study the {\it quantum} mechanics of this system 
by the Euclidean path integral over $q(\tau)$ and $x_{\alpha}(\tau)$, 
 \begin{eqnarray}
Z&=&\int{\cal D}q~\hbox{$ \displaystyle\prod_{\alpha}$}
\int{\cal D}x_{\alpha}~
e^{-S_{\rm E}},\label{(2a)}\\
{S}_{\rm E}[~q,\{x_{\alpha}\}]
&=&
\int d{\tau}~\left\{
\frac{1}{2}M{\dot q}^{2} + V_0(q)
\right.
\nonumber\\
+\sum_{\alpha}&&\!\!\!\!\!\!\!\!\!
\left.\left[~\frac{1}{2}m_{\alpha}{\dot x_{\alpha}}^{2}
+\frac{1}{2}m_{\alpha}\omega_{\alpha}^{2}x_{\alpha}^{2}
+q C_{\alpha}x_{\alpha}\right]
\right\}.
\label{(2b)}
 \end{eqnarray}
We integrate the variable $x_{\alpha}$
to define the effective action for the target system only, 
 \begin{eqnarray}
Z=\int{\cal D}q\!\!&\exp&\!\! \left\{-\!\int \!d\tau\!\!
\left[\frac{1}{2}M{\dot q}^{2}+V_0(q)\right]\!\!
-\Delta S[q]\right\},\label{(2c)}\\
\Delta S[q]\!\!\!&=&\!\!\!
-\ln 
\hbox{$ \displaystyle\prod_{\alpha}$}\int{\cal D}x_{\alpha}~
\!\!\!\exp\left\{-\!\int \!d\tau
\left[\frac{1}{2}m_{\alpha}{\dot x_{\alpha}}^{2}
\right.\right.
\nonumber\\
&&+\left.\left.\frac{1}{2}m_{\alpha}\omega_{\alpha}^{2}x_{\alpha}^{2}
+q C_{\alpha}x_{\alpha}\right]\right\}
,\label{(2d)}
 \end{eqnarray}
where $\Delta S[q]$ is the effective interaction term 
generated by the quantum fluctuation 
of the environment $x_{\alpha}$.
\par
Caldeira and Leggett studied the influence of $\Delta S$ on 
the quantum tunneling of $q$ and they found that the Ohmic 
dissipation suppressed the quantum tunneling \cite{cl1}.
Much theoretical work has been done following their analysis
\cite{cl2, fiss1, fiss2, cha, bm, fz, weiss},
and the validity of their results has been suggested 
by elaborate experiments \cite{fvdz}.
However, the theoretical works so far 
(instanton, perturbation, etc.) depend on 
some expansions with respect to small parameters, 
which are not always valid. 
\par
We adopt the nonperturbative renormalization-group (NPRG) method
\cite{wk, wh, ap, morris, aoki1, aoki2, ahtt, kt, za}
to analyze the model.
The formulation of the NPRG 
does not require any series expansion.  
The NPRG method also helps us to interpret the effects of $\Delta S$
as effective infrared or ultraviolet cutoffs for the quantum
fluctuation of the target system variable, 
and we readily understand 
the dissipation effects as suppression or enhancement of the
quantum features of the effective target system. 
Furthermore, the NPRG method is in general
very powerful for studying the phase structure 
and critical phenomena.
In dissipative quantum mechanics, a localization
(quantum-classical) transition has been suggested to occur, 
and it is investigated by the NPRG method 
in a straightforward way 
to estimate the divergence of the localization susceptibility.
We reported a NPRG analysis for Ohmic dissipation 
in a brief form \cite{ah}.
In this article we describe the method in detail and treat general
types of dissipation.
\par
This paper is organized as follows. 
In Sec. II we illustrate classical and quantum features of
the Caldeira-Leggett model.
In Sec. III we present our strategy to apply the NPRG to 
quantum dissipative systems.  
The NPRG equation for dissipative quantum mechanics is derived 
and the effect of $\Delta S$ is analyzed.
In Sec. IV we solve the NPRG equation 
in a double well potential system and calculate the localization 
susceptibility to evaluate the critical dissipation for the 
quantum-classical transition.
Sec. V contains a summary and discussion.

\section{The Caldeira-Leggett model}
\subsection{Classical equation of motion}

We start with the classical features of the model and show that 
the dissipative classical equation of motion
is derived from the action (\ref{(1)}).
By eliminating the environmental variable $x_{\alpha}$
with the boundary condition $x_{\alpha}(t=-\infty)=0$, 
the equation of motion of $q$ takes the following form:
\begin{eqnarray}
M \ddot{q}=-\frac{\partial V_{0}}{\partial q} 
&-& \int^{\infty}_{-\infty}
\frac{d\omega}{2\pi}~e^{-i\omega t}\tilde{q}(\omega)\nonumber\\
&\times&
\sum_{\alpha}
\frac{C_{\alpha}^2}
{m_{\alpha}}
\frac{1}
{~(\omega+i\epsilon)^2-\omega_{\alpha}^2} ,\nonumber\\
=
-\frac{\partial V_{0}}{\partial q} 
&-& \int^{\infty}_{-\infty}
\frac{d\omega}{2\pi}~e^{-i\omega t}\tilde{q}(\omega)\nonumber\\
&\times&\int^{\infty}_{0}
\frac{d\omega^{\prime}}{2\pi}
J(\omega^{\prime})
\frac{4 \omega^{\prime}}{~(\omega+i\epsilon)^2-\omega^{\prime 2}} ,
\label{(4)}
 \end{eqnarray}
where $\epsilon$ is an infinitesimal positive regulator
to assure the boundary condition.
We have introduced the spectral density function $J(\omega)$ 
which characterizes the environment,
\begin{eqnarray}
J(\omega)&=&\sum_{\alpha}
~\frac{C^2_{\alpha}}{4 m_{\alpha}\omega_{\alpha}}
~(2\pi)~\delta(\omega-\omega_{\alpha}).\label{(5)}
\end{eqnarray}
A suitable choice of $J(\omega)$ generates dissipation. 
For example, if we set $J(\omega)=\eta~\!\omega$, 
then Eq. (\ref{(4)}) reads
 \begin{eqnarray}
M \ddot{q}
&=& -\frac{\partial}{\partial q} 
\left(V_{0}-\frac{\eta\omega_c}{\pi}q^2\right)
- \eta ~\!\dot{q} ,
\label{(6)}
 \end{eqnarray}
where $\omega_c$ is an ultraviolet cutoff of the environmental 
oscillator frequencies. 
The effects of the environment $x_{\alpha}$ appear as
the Ohmic dissipation term $- \eta ~\!\dot{q}~$ and 
the correction to the potential $-(\eta\omega_c/\pi)q^2$. 
The latter can be absorbed 
by a local counterterm and hereafter we are interested only in
the dissipation effect.
\par
We study general types of dissipation obtained by
$J(\omega)=\eta~\!\omega^s~~(s=1,3,5,...)$,
where cases of $s>1$ are called super-Ohmic. 
Then the effective equation of motion is 
 \begin{eqnarray}
M \ddot{q}
&=& -\frac{\partial}{\partial q} 
V_{0}
-(-1)^{(s-1)/2}~\eta~\frac{d^{s}q}{dt^{s}},
\label{(7)}
 \end{eqnarray}
where only the highest derivative term 
breaking the time reversal symmetry has been included.
The origin of the time reversal symmetry breaking
is the boundary condition 
$x_{\alpha}(t=-\infty)=0$ 
for eliminating the variable $x_{\alpha}$.  
Note that, in order to get a function $J(\omega)$
that is exactly a simple polynomial in $\omega$,
an infinite number of gapless oscillators $x_{\alpha}$
are required.
This infinite number of oscillators is actually supplied 
by some field degrees of freedom 
(the electromagnetic field, the phonon field, etc.). 

\subsection{Dissipation term in the effective action}

In the Euclidean path integral formulation of quantum mechanics, 
the correction term $\Delta S$ due to the environment
is defined in Eq. (\ref{(2d)}),
\begin{eqnarray}
\Delta S[q]&=&-~\int d\tau\!\int ds~q(\tau)
~\alpha(\tau -s)~q(s), \label{(8)}
\end{eqnarray}
where the nonlocal coupling coefficient is given by
\begin{eqnarray}
\alpha(\tau -s)&=&
\int \frac{dE}{2\pi}
~\sum_{\alpha}~\frac{C^2_{\alpha}}{2 m_{\alpha}}
\frac{1}{E^2+\omega^2_{\alpha}}~e^{iE(\tau -s)}\nonumber\\
&=&\sum_{\alpha}
~\frac{C^2_{\alpha}}{2 m_{\alpha}}
\frac{1}{2\omega_{\alpha}}e^{-\omega_{\alpha}|\tau -s|}\nonumber\\
&=&\!\int^{\infty}_{0} 
\frac{d\omega}{2\pi}~J(\omega)~e^{-\omega|\tau -s|}.\label{(11)}
\end{eqnarray}
Generally $\Delta S$ consists of 
the local component $\Delta S_{\rm L}$ and 
the genuine nonlocal component $\Delta S_{\rm NL}$.
We identify $\Delta S_{\rm NL}$ component as the dissipation term,
since it corresponds to the odd power term in the Fourier transform,
and thus it is related to time reversal symmetry breaking.
On the other hand, we do not consider the $\Delta S_{\rm L}$ component 
supposing that they are renormalized to vanish with 
suitable counterterms.
In fact they are all even power terms in the Fourier transform and
are irrelevant to dissipation.
\par
For example, in the case of Ohmic dissipation $J(\omega)=\eta~\!\omega$,
we identify the dissipative part as follows:
\begin{eqnarray}
\Delta S_{\rm L}[q]&=&-~\!\tilde{\alpha}(E=0)
\int d\tau~q^2(\tau) ,\label{(12)}\\
\Delta S_{\rm NL}[q]&=&\!\!\frac{1}{2}\int\! d\tau\!\int\! ds~\!
\left[q(\tau)-q(s)\right]^2
\alpha(\tau -s) ,\label{(13)}
 \end{eqnarray}
where 
\begin{eqnarray}
\tilde{\alpha}(E=0)&=&
\sum_{\alpha}~\frac{C^2_{\alpha}}{2 m_{\alpha}}
\frac{1}{\omega^2_{\alpha}}\nonumber\\
&=&\!\int^{\infty}_{0} 
\frac{d\omega}{2\pi}~J(\omega)\frac{2}{\omega}=\frac{\eta}{\pi}\omega_c ,
\label{(14)}\\
\alpha (\tau -s)&=&
\!\int^{\infty}_{0} 
\frac{d\omega}{2\pi}~\eta~\omega~e^{-\omega|\tau -s|}\nonumber\\
&=&\frac{\eta}{2\pi}\frac{1}{|\tau -s|^2} .\label{(15)}
\end{eqnarray}
If we take the ultraviolet cutoff to be infinite,
the local component $\Delta S_{\rm L}$ diverges.
We prepare the counterterm 
$\Delta S _{\rm CT}=-\Delta S_{\rm L}=
\int d\tau (\eta \omega_c/\pi)q^2$ 
to remove it. 
This procedure is nothing but 
the standard mass renormalization \cite{comment2}. 
The dissipation term for the Ohmic dissipation is given by
\begin{eqnarray}
\Delta S_{\rm NL}[q]&=&\!\frac{\eta}{4\pi}\int d\tau\!\int ds~
\frac{\left[~q(\tau)-q(s)~\right]^2}{|\tau -s|^2},\label{(16)}
 \end{eqnarray}
which is considerably nonlocal. However, note that 
this term never breaks time reversal symmetry.
When evaluating the Euclidean path integral
we do not use any specific boundary condition 
for the integrated variables $x_{\alpha}$.
Therefore we should be careful when we call $\Delta S_{\rm NL}$
the ``dissipation term.'' 
It means that its classical effect is dissipative 
when the retarded boundary condition is employed.
\par
The Fourier transform of the Ohmic dissipation term 
$\Delta S_{\rm NL}$ reads
\begin{eqnarray}
\Delta S_{\rm NL}[q]
&=&-\int \frac{dE}{2\pi}~\tilde{q}~\!(E)~\!\tilde{q}~\!(-E)\nonumber\\
&&~~~\times
\sum_{\alpha}
\frac{C^2_{\alpha}}{2 m_{\alpha}}\left(
\frac{1}{E^2+\omega_{\alpha}^2}-\frac{1}{\omega_{\alpha}^2}
\right)\nonumber\\
&=&-\int \frac{dE}{2\pi}~\tilde{q}~\!(E)~\!\tilde{q}~\!(-E)\nonumber\\
&&~~~\times
\int_{0}^{\infty}\frac{d\omega}{2\pi}~J(\omega)~2\left(
\frac{\omega}{E^2+\omega^2}-\frac{1}{\omega}
\right)\nonumber\\
&=&
\int \frac{dE}{2\pi}~\tilde{q}~\!(E)~\!\tilde{q}~\!(-E)\nonumber\\
&&~~~\times
~\frac{\eta}{\pi}E^2\int^{\infty}_{0}d\omega
\frac{1}{\omega^2+E^2}\nonumber\\
&=&\!\frac{1}{2}~\!\!\int \frac{dE}{2\pi}
~\eta~|E|~
\tilde{q}~\!(E)~\!\tilde{q}~\!(-E).\label{(20)}
 \end{eqnarray}
As a typical form of the ``dissipation term,'' 
the absolute value of $E$, $|E|$, appears.
For a general environment $J(\omega)=\eta\omega^{s}$, 
it is expressed as 
\begin{eqnarray}
\Delta S_{\rm NL}[q]\!\!
&=&\!\!\frac{1}{2}~\!\!\int \frac{dE}{2\pi}~\!
(-1)^{(s-1)/2}\eta~\!|E|^s~\!
\tilde{q}~\!(E)~\!\tilde{q}~\!(-E).
\nonumber\\
&&\label{(21)}
 \end{eqnarray}
Note that $\eta$ is a dimensionful parameter, 
[$\eta$]$=$[$E^{2-s}$]. 

\section{NPRG approach for quantum dissipative systems}
\subsection{Our strategy}

Now let us proceed to the NPRG analysis 
of the Caldeira-Leggett model \cite{comment1}.
We study the influence of $\Delta S_{\rm NL}$ on 
the quantum behaviors of $q$. 
Note that $\Delta S_{\rm NL}$ is symmetric under time reversal,
and we will not see any real dissipating phenomena.
As shown below, the effect of $\Delta S_{\rm NL}$ modifies the 
propagator of $q$, and is interpreted as various cutoff effects 
in the NPRG point of view. 
Using the NPRG method we evaluate 
the effective potential (free energy)
of the expectation value of the target system variable $q$,
and calculate the localization 
susceptibility.
We will evaluate the critical dissipation 
for the localization phase transition using 
the critical scaling behavior.

\subsection{NPRG analysis of ordinary quantum mechanics}

Originally the NPRG method was formulated and used mainly 
in statistical mechanics 
or quantum field theory to analyze critical phenomena
\cite{wk, wh, ap, morris, aoki1, aoki2}.
Recently, the NPRG method has even been found effective 
in quantum mechanics, 
particularly for nonperturbative analysis
\cite{ahtt, kt, za}.
The formulation of the NPRG method for standard ($\eta=0$) 
quantum mechanics is summarized as follows.
\par
In the NPRG method, the theory is defined by 
the Wilsonian effective action $S_{\Lambda}[q]$.
This is an effective theory with 
an ultraviolet energy cutoff $\Lambda$.
Then we carry out a path integration over the highest energy
degrees of freedom of $\tilde{q}(E)$, 
$\Lambda-\Delta \Lambda<|E|\le\Lambda$,
called the ``shell mode.''
After the shell mode integration, we get 
a new effective theory denoted by 
$S_{\Lambda-\Delta\Lambda}[q]$, 
which is a functional of lower energy modes only.
This procedure is called the renormalization transformation and
it defines an iterative sequence of points in the theory space 
spanned by all possible functionals $S[q]$.
By taking the continuous transformation limit $\Delta \Lambda\to 0$, 
we reach a (functional) differential equation called the NPRG
equation, 
\begin{eqnarray}
\Lambda\frac{\partial S_{\Lambda}[q]}{\partial \Lambda}
&=&\beta \left[ S_{\Lambda}[q] \right],\label{(21a)}
 \end{eqnarray}
where the $\beta$ functional is explicitly given 
in the literature \cite{aoki1,aoki2}.
This equation is equivalent to a set of 
infinite-dimensional coupled ordinary 
differential equations, and defines a flow line in the theory space.
When we integrate the equation to the infrared limit ($\Lambda=0$), 
we arrive at $S_{\Lambda=0}[q]$ 
which has the full quantum information of the theory.
\par
Although the $\beta$ functional is obtained without any
approximations, we have to make some approximations 
to solve the NPRG equation.
We define a sub theory space in the full theory space and project the
full equation (\ref{(21a)}) onto the subspace, which defines a reduced 
equation on the subspace. 
As a subspace, we take the local potential space spanned by actions of
the following form: 
\begin{eqnarray}
S_{\Lambda}[q]&=&\int d\tau \left[
\frac{1}{2}~\dot{q}^2 + V_{\Lambda}(q)
\right].\label{(21b)}
 \end{eqnarray}
In this subspace, the potential function $V_{\Lambda}(q)$ 
defines a theory, and our subspace is 
still an infinite-dimensional function space.  
Then the reduced renormalization transformation is given by 
\begin{eqnarray}
V_{\Lambda-{\mit\Delta}\Lambda}(q)&=&V_{\Lambda}(q)\nonumber\\
&+&\frac{1}{2}~{\int_{\rm shell}}\frac{dE}{2\pi}~{\ln}\left(
E^2
+\frac{\partial^2 V_{\Lambda}}{\partial q^2}
\right),\label{(22)}
 \end{eqnarray}
where we path integrate the shell mode up to one-loop order.
Taking the continuous limit $\Delta\Lambda\to 0$, the one-loop
evaluation becomes exact, and we obtain the reduced NPRG equation
 \begin{eqnarray}
 \Lambda\frac{\partial V_{\Lambda}}{\partial \Lambda}&=&
-~\!\frac{1}{2\pi}~\!\Lambda
  ~\!\ln\left(1+
\frac{1}{\Lambda ^2}\frac{\partial ^2
  V_{\Lambda}}{\partial q ^2} \right).\label{(23)}
 \end{eqnarray}
We will work with this equation, which is called the 
local potential approximated 
Wegner-Houghton (LPA W-H) equation
\cite{wh, ap, morris, aoki1, aoki2}.
Note that no perturbative expansion has been adopted 
to obtain Eq. (\ref{(23)}).
We may improve the approximation by enlarging the sub theory space,
and it is completely free of any sort of series expansion and
nonconvergence problems.
Due to these features, we expect that the NPRG method will work 
excellently for the nonperturbative analysis of quantum systems.
\par
The target equation (\ref{(23)})
is a partial differential equation of 
${V}_{\Lambda}$ with respect to $q$ and $\Lambda$.
We solve it by lowering $\Lambda$ 
from the initial cutoff $\Lambda _0$ 
where the initial potential $V_{\Lambda_0}$ is given 
by the potential term in the original quantum action, that is,
$V_{\Lambda_0}(q)=V_0(q)$.
At the infrared limit, we get the effective potential 
$V_{\rm eff}(q)=\lim_{\Lambda\to 0}V_{\Lambda}(q)$, from which 
physical quantities are evaluated.
First, the expectation value of $q$ in the ground state 
$\langle q\rangle$ is
determined by the stationary condition,
 \begin{eqnarray}
\left.\frac{d V_{\rm eff}}{d q}\right|
_{q=\langle q\rangle}=0,
 \end{eqnarray}
and then the ground state energy is given by
 \begin{eqnarray}
E_0=V_{\rm eff}(q=\langle q\rangle).
 \end{eqnarray}
Also we obtain 
the energy gap of the system, 
 \begin{eqnarray}
\Delta E&\equiv&E_1-E_0=m_{\rm eff}\equiv\sqrt{\left.\frac{d^2 V_{\rm eff}}
{d q^2}\right|}_{q=\langle q\rangle},\label{(24)}
 \end{eqnarray}
where $m_{\rm eff}^2$ 
is nothing but the curvature of the potential at the minimum
\cite{comment2}.
\par
As the initial potential $V_{\Lambda_0}$, 
we choose the double well type,
$V_{\Lambda_0}=-\frac{1}{2}m_0^2 q^2+\lambda_0 q^4$,
throughout this article.  
We work with the units $\hbar=M=1$ and measure 
all dimensionful quantities by $m_0$, that is,  
$\Lambda[m_0]$, $V_{\Lambda}[m_0]$, $q[m_0^{-1/2}]$, 
$\lambda_0[m_0^3]$, etc.  
Hereafter we will omit the units from the numerical values.

\subsection{ NPRG equation with dissipation}

We derive the NPRG equation for quantum mechanics 
with a dissipation term.
We employ the local potential approximation as above.
Then only the potential part of the action is changed by the 
renormalization transformation.
We integrate out the shell mode 
up to one-loop order, 
 \begin{eqnarray}
V_{\Lambda-{\mit\Delta}\Lambda}(q)&=&V_{\Lambda}(q)
+\frac{1}{2}~{\int_{\rm shell}}\frac{dE}{2\pi}\nonumber\\
\times&&\!\!\!\!\!\!\!\!\!
{\ln}\left(E^2+
(-1)^{(s-1)/2}\eta|E|^s
+\frac{\partial^2 V_{\Lambda}}{\partial q^2}
\right), \label{(25)}
 \end{eqnarray}
where we have used the modified propagator due to the $\eta$ term.
Taking the limit ${\mit\Delta}\Lambda\to 0$,
we have the NPRG equation
\begin{eqnarray}
 \Lambda\frac{\partial V_{\Lambda}}{\partial\Lambda}&=&
-~\!\frac{1}{2\pi}~\!\Lambda\nonumber\\
\times&&\!\!\!\!\!\!\!\!\!
\ln\left(1+(-1)^{(s-1)/2}
\eta \Lambda ^{s-2}
+\frac{1}{\Lambda ^2}\frac{\partial ^2
  V_{\Lambda}}{\partial q ^2} \right).\label{(26)}
 \end{eqnarray}
This should be called the LPA Wegner-Houghton equation with dissipation; 
it has an extra term 
$(-1)^{(s-1)/2}
\eta \Lambda ^{s-2}~$
caused by the dissipation term.

\subsection{Dissipation effects as effective cutoffs}

We evaluate the effects of the dissipation term 
on the quantum behaviors of $q$ by solving the NPRG equation  
(\ref{(26)}).
Before proceeding to the numerical calculation,
let us pay attention to the form of the propagator
modified by the dissipation term
and discuss the dissipation effect qualitatively.
The inverse propagator $\Delta^{-1}(E)$ consists of three parts,   
the original part of the kinetic term $S_{\rm kin}$ and 
the mass term $S_{\rm mass}$, and 
the dissipation part $\Delta S_{\rm NL}$,
 \begin{eqnarray}
&&S_{\rm kin}[q]+S_{\rm mass}[q]+\Delta S_{\rm NL}[q]\nonumber\\
&&=\!\frac{1}{2}
\int \!\frac{dE}{2\pi}\!
\left[
E^2\!+\!\frac{\partial^2 V_{\Lambda}}{\partial q^2}\!+\!
(-1)^{(s-1)/2}\eta~\!|E|^s
\right]
\tilde{q}~\!(E)~\!\tilde{q}~\!(-E)\nonumber\\
&&=\!\frac{1}{2}\int \frac{dE}{2\pi}~
\Delta^{-1}(E)~
\tilde{q}~\!(E)~\!\tilde{q}~\!(-E).
 \end{eqnarray}
For the high energy region, the mass term can be neglected.
For the low energy region, the mass term dominates and the NPRG 
equation effectively reduces to a free form without interactions.
Therefore when we qualitatively discuss the dissipation effects
we can ignore the mass term.
\par
\begin{figure}
\includegraphics[width=80mm,height=80mm]{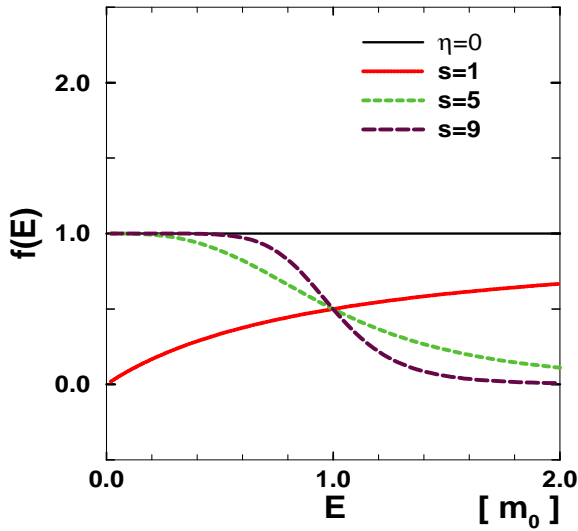}
\caption{\label{fig:fp1} Cutoff factor $f(E)$ for $s=1+4n$ with 
$\eta=1[m_0^{2-s}]$.}
\end{figure}
\begin{figure}
\includegraphics[width=80mm,height=80mm]{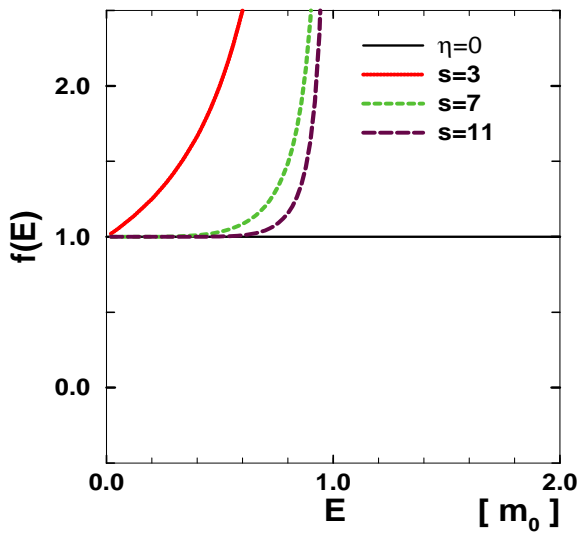}
\caption{\label{fig:fp2} Cutoff factor $f(E)$ for $s=3+4n$ with 
$\eta=1[m_0^{2-s}]$.}
\end{figure}

We compare the propagator with the standard form without dissipation,
 \begin{eqnarray}
f(E)&\equiv& E^2 \Delta (E)\nonumber\\
&=&\frac{E^2}{E^2+(-1)^{(s-1)/2}\eta~\!|E|^s}.
 \end{eqnarray}
Some typical forms of $f(E)$ are shown in
Fig. \ref{fig:fp1} and Fig. \ref{fig:fp2} with $\eta =1$. 
The effects of the dissipation term $\Delta S_{\rm NL}$
can be interpreted as {\it cutoff effects} and 
are classified into three groups.

For $s=1$ Ohmic dissipation, since $f(E)\to 0$ as $E\to 0$ and $f(E)\to 1$ 
as $E\to \infty$, the dissipation term $\Delta S_{\rm NL}$
suppresses the propagation
in the $E\to 0$ region, that is, it is equivalent to an 
{\it effective infrared cutoff}.
On the other hand, for $s=1+4n$ super-Ohmic dissipation,
the propagation is suppressed in the $E\to \infty$ region, 
and the dissipation term $\Delta S_{\rm NL}$ is equivalent to an 
{\it effective ultraviolet cutoff}.
These cutoffs have their own typical energy scales 
$E_c={\eta}^{-1/(s-2)}$
where $f(E)=\frac{1}{2}$. 
\par
Next, we consider the cases of $s=3+4n$ super-Ohmic dissipation 
shown in Fig. \ref{fig:fp2}.
In these cases $f(E)$ shows singular behavior in that 
it diverges at the energy scale $E_c={\eta}^{-1/(s-2)}$.
This means that the theory is normal only below
the ultraviolet cutoff energy $E_c$.
In the normal region $f(E)$ is larger than $1$
and the propagation is enhanced.
\par
As discussed above,
$\Delta S_{\rm NL}$ acts as  
various types of effective cutoffs, which suppress or
enhance the quantum fluctuation in energy dependent ways.
\par
The NPRG method we employ now is a method to
divide the variables $q(\tau)$ into energy shell modes, 
and to evaluate the quantum fluctuation of each shell mode 
systematically from the higher energy modes.
It is suitable for studying the 
energy dependent cutoff effect appearing 
in dissipative quantum mechanics.
Through the NPRG analysis of ordinary ($\eta=0$) quantum mechanics,  
we know that  
the effective energy region where the quantum fluctuations are
significant depends on the mass scale of the theory \cite{ahtt}.
This fact implies that, if the cutoff effect originating from 
the dissipative term $\Delta S_{\rm NL}$ 
is large in this effective energy region,
it may modify the 
phase structure of the system.
In this way, the NPRG method allows us to understand intuitively the 
effect of $\Delta S_{\rm NL}$ as an effective energy dependent cutoff
and discuss its strength by comparing it with the effective energy region
of the original system. 
Such a viewpoint is a unique advantage of the NPRG approach.

\section{Analysis of Dissipative Quantum Tunneling}
\subsection{Macroscopic quantum coherence}

The Caldeira-Leggett model consists of 
two classes of variables,
the target system $q$ and 
the environmental degrees of freedom $\{x_{\alpha}\}$.
We regard the system variable $q$ 
as a {\it macroscopic} collective coordinate and 
discuss how the quantum mechanical behavior of the 
macroscopic  coordinate $q$ 
is affected by the environmental effects.
As a typical phenomenon, the quantum tunneling of $q$
has been analyzed 
by many authors \cite{cl1,cl2, fiss1, fiss2, cha, bm}.
We set the bare potential of $q$ as 
the double well type, $V_0(q)=-\frac{1}{2}q^2+\lambda _0 q^4$.
In this system with vanishing $\eta$ the $q$ mode tunnels 
through the potential barrier and oscillates 
between the two wells.
The main question is whether or not 
it also oscillates even in the
$\eta\ne 0$ dissipative system. 
Such an oscillatory behavior is called 
{\it macroscopic quantum coherence} and 
the inclination to cease the oscillation  
corresponds to {\it decoherence} \cite{weiss}.  
\par
Now we briefly summarize the previous results    
obtained for macroscopic quantum coherence in double well
potential systems with various types of dissipation 
$J(\omega)=\eta~\!\omega^s$.
Usually the first energy gap $\Delta E=E_1-E_0$ is calculated 
as the physical quantity because it corresponds to the 
tunneling amplitude between the two wells.
For $s=1$ Ohmic dissipation, Caldeira and Leggett evaluated it
by using the semi-classical (instanton) approximation, which is 
valid for $\lambda _0\to 0$ and perturbation with 
respect to $\eta$, and they found that the dissipation   
$\Delta S$ suppresses the tunneling \cite{cl1,cl2}.
Renormalization-group analyses have been done for the 
instanton gas system within the dilute gas approximation, 
which is also valid for 
$\lambda _0\to 0$ and $\eta\to 0$ \cite{cha, bm}.
They predict the remarkable phenomenon that     
$\Delta E$ vanishes
at a critical value $\eta_c=2\pi\lambda_0$, 
where decoherence occurs.
This is often called the {\it quantum-classical transition} 
because the system looks classical in the limited sense 
that quantum tunneling does not occur.  
This is nothing but the spontaneous $Z_2$ symmetry breaking. 
The energy gap $\Delta E$ plays an important role 
as the order parameter of the transition.
\par
On the other hand, for $s=3$ super-Ohmic dissipation, 
the possible enhancement of $\Delta E$ is shown by 
perturbative calculus with respect to $\eta$ 
in the canonical formulation \cite{fiss1,fiss2}. 
This result is interpreted as the effect of the higher excited states 
($E_n, n\ge 2$), and it cannot be reproduced 
by the standard instanton calculus.
For the cases of $s=5,7,9,...$ super-Ohmic dissipation,
no detailed analyses have been carried out. 
The dissipation effects in these cases 
are expected to be weak by simple power counting,
since they are irrelevant in the infrared region.      
\par
These results we listed above are obtained by using the semiclassical 
approximation and/or perturbation theory with respect to $\eta$.
Their reliability depend on the 
smallness of the couplings $\lambda_0$ and $\eta$. 
Therefore, to get more general and reliable results 
free of this limitation,
we must employ an analyzing tool which does not 
need a series expansion with respect
to any couplings.
We expect our method of using the NPRG
will work best, particularly for the large coupling region.

\subsection{NPRG analysis}

\begin{figure}
\includegraphics[width=80mm,height=80mm]{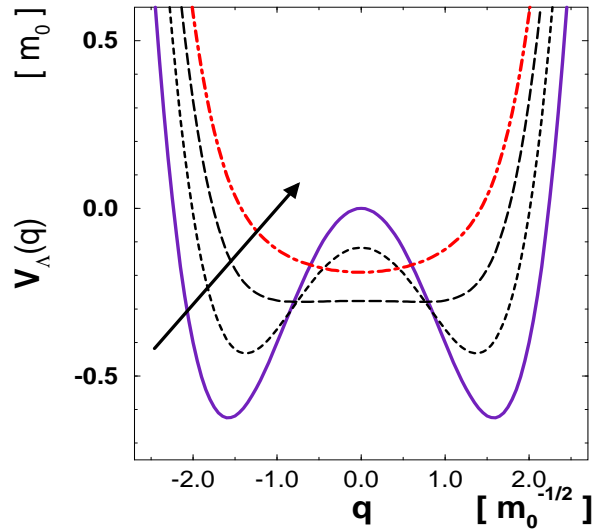}
\caption{\label{fig:epote} Flow of the potential for 
$\lambda_0=0.1[m_0^3]$.}
\end{figure}

We analyze dissipative quantum tunneling 
by solving the NPRG equation (\ref{(26)}) numerically.
This equation is a two-dimensional partial differential equation 
for the Wilsonian effective potential $V_{\Lambda}(q)$ 
with respect to $\Lambda$ and $q$. 
The initial condition of the potential
is the bare potential $V_{\Lambda_{0}}(q)=-\frac{1}{2}q^2+\lambda _0 q^4$ 
at the initial cutoff $\Lambda_0$.
We solve the differential equation in the infrared limit 
$\Lambda\to 0$ and obtain the physical effective potential 
$V_{\rm eff}(q)$.
The typical change of the potential toward the infrared is
shown in Fig. \ref{fig:epote}. 
We see two basic features of the flow.
The potential at the origin moves up, which finally gives 
the ground state energy of the zero point oscillation.
The second derivative of the potential at the origin also 
increases from negative to positive, changing the potential form 
from a double well to a single well.

\begin{figure}
\includegraphics[width=80mm,height=80mm]{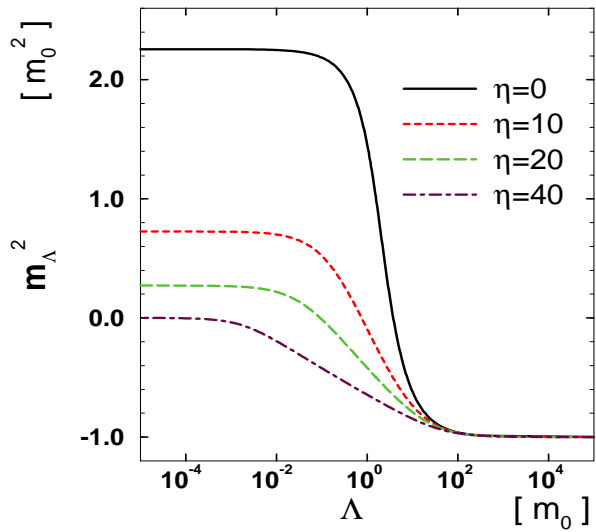}
\caption{\label{fig:Dmrun} Running of $m^2_{\Lambda}$ for
$\lambda_0=1.0[m_0^3]$. All values of $\eta$ are
measured in units of $[m_0]$. }
\end{figure}

We note here the actual treatment of the infrared and the ultraviolet 
limit in the numerical calculation.
The Wilsonian effective potential changes toward the infrared and 
finally stops changing at some finite scale of $\Lambda$, 
below which the quantum fluctuations are suppressed due to the 
mass of the system.
Therefore before lowering the scale $\Lambda$ to vanishing, 
we get the infrared limit results.
As for the initial cutoff, it should be infinite 
since the target system is quantum mechanics 
without an ultraviolet energy cutoff.
We start with a large value of the initial cutoff 
to solve the equation,
and check that the physical results given by the infrared potential 
change little when we move the initial cutoff up further.
Then we regard our numerical results as those without any cutoff.
\par
We can exploit the physical information of the quantum system 
from $V_{\rm eff}(q)$.
First of all, the mass squared starting with a negative value  
$(-1)$ finally reaches a positive value (for small $\eta$), 
$m_{\rm eff}^{2}>0$, that is, the effective potential $V_{\rm eff}(q)$ 
becomes a single well form.
The running of the mass squared is plotted in Fig. \ref{fig:Dmrun} 
for $\lambda_0=1.0$. 
This means that, due to the quantum tunneling, the $Z_2$ symmetry
(parity) does not break spontaneously, 
and the expectation value of $q$ vanishes.
The tunneling effects are automatically incorporated 
when we integrate (solve) the NPRG equation toward the infrared.
\par
When the dissipation effects are included, the quantum propagator may
be suppressed, and then such running of $m^2_{\Lambda}$ 
as shown in Fig. \ref{fig:Dmrun} is also suppressed.
If the running is not enough, the mass squared may not become positive 
even at the infrared limit.
This indicates spontaneous symmetry breaking, and localization
should be observed.
\par
For nonvanishing $\eta$, the correspondence between 
the energy gap $\Delta E$ and the effective mass $m_{\rm eff}$  
Eq. (\ref{(24)}) does not hold.   
The long range ($\tau\to\infty$) behavior of the two point function 
 \begin{eqnarray}
\lim_{\tau\to\infty}
\int \frac{dE}{2\pi}e^{iE\tau}\frac{1}{E^2+\eta|E|+m^2_{\rm eff}}
 \end{eqnarray}
is actually independent of $m_{\rm eff}$ and is determined by 
$\eta$ with power damping behavior 
even in the case of infinitesimal $\eta$.
This singular behavior comes from the nature of the environmental 
degrees of freedom where the harmonic oscillator frequencies are
assumed to be distributed continuously up to zero ($\omega_{\alpha}=0$). 
Even if there is an infrared cutoff, it does not change the situation 
much and the lowest frequency environment dominates the long range 
correlator of the target system. 
In this sense semiclassical arguments depending on 
the long range transition amplitude seem doubtful
for the purpose of investigating the localization transition. 
\par
Here, instead of analyzing the long range correlator, we adopt 
another physical quantity to describe the possible quantum-classical 
transition, the localization susceptibility. 
We suppose that a source term $J$ that couples to the zero mode of 
$\tilde{q}(E)$,
 \begin{eqnarray}
\tilde{q}(0)=\int^{T/2}_{-T/2}q(\tau)d\tau, 
 \end{eqnarray}
is included in the quantum action.
Then the localization susceptibility $\chi$ is defined by
 \begin{eqnarray}
\chi\equiv
\frac{1}{T}
\left.\frac{d \langle \tilde{q}(0) \rangle_{J}}{d J}\right|_{J=0}
=\frac{1}{T}
\left.\frac{d^2 \ln Z}{d J^2}\right|_{J=0},
 \end{eqnarray}
and it is exactly given by the effective potential as follows:
 \begin{eqnarray}
\chi=\left(\frac{d^2 V_{\rm eff}(q)}{d q^2}\right)^{-1}=
\frac{1}{m^2_{\rm eff}}.
 \end{eqnarray}
These definitions are valid also for nonvanishing $\eta$.
If we find a divergent and scaled behavior of $\chi$
toward $\eta=\eta_c$, we may conclude that 
the localization transition occurs at $\eta_c$ 
and it is a second order phase transition.

\begin{figure}
\includegraphics[width=80mm,height=80mm]{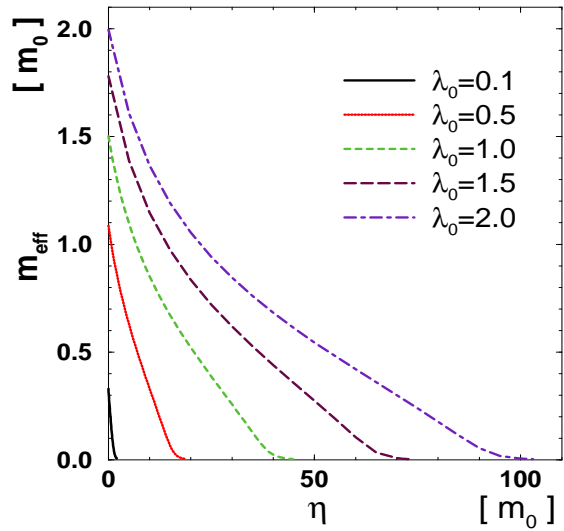}
\caption{\label{fig:ohm} The effective mass $m_{\rm eff}$ for $s=1$ 
with initial cutoff $\Lambda_0=10^4[m_0]$. 
All values of $\lambda_0$ are
measured by $[m_0^3]$ unit.}
\end{figure}

\begin{figure}
\includegraphics[width=80mm,height=80mm]{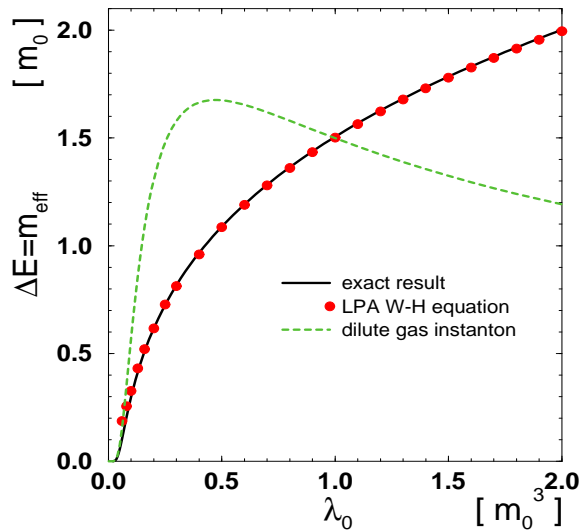}
\caption{\label{fig:gap} Energy gap estimates without dissipation.}
\end{figure}

The results for $s=1$ Ohmic dissipation
are shown in Fig. \ref{fig:ohm} for several $\lambda_0$.
In previous works \cite{ahtt,kt,za} for the $\eta=0$ system,
the NPRG analyses have been found to work excellently 
in these $\lambda _0$ regions
while the dilute gas instanton approximation 
does not work at all there.
In Fig. \ref{fig:gap} we recapitulate the results
for the energy gap with $\eta=0$ 
where those by the dilute gas instanton are also plotted for
comparison and the exact values are calculated by solving
the Schr\"odinger equation.
We see that for the large $\lambda_0$ region ($\lambda_0\gsim 0.1$) 
the NPRG results looks excellent while the instanton method 
is out of its effective range just as is expected.
Therefore we expect that the NPRG analysis also works well 
for these $\lambda _0$ values
even in $\eta\ne 0$ systems.

\begin{figure}
\includegraphics[width=80mm,height=80mm]{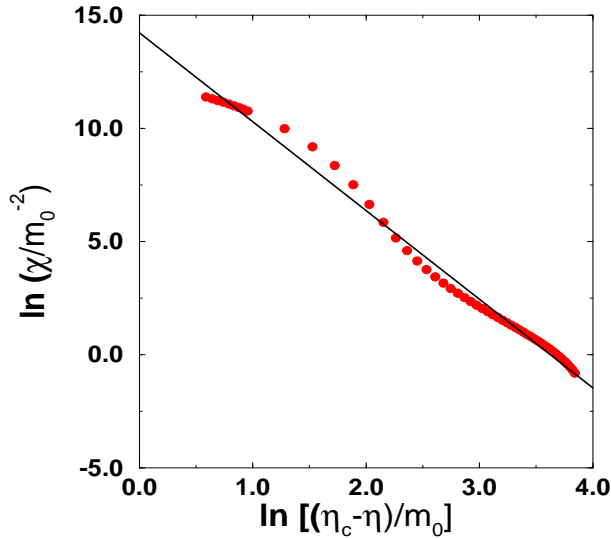}
\caption{\label{fig:sus} Critical scaling fit of 
localization susceptibility $\chi$ for 
$\lambda_0=1.0[m_0^3]$. }
\end{figure}

\begin{figure}
\includegraphics[width=80mm,height=80mm]{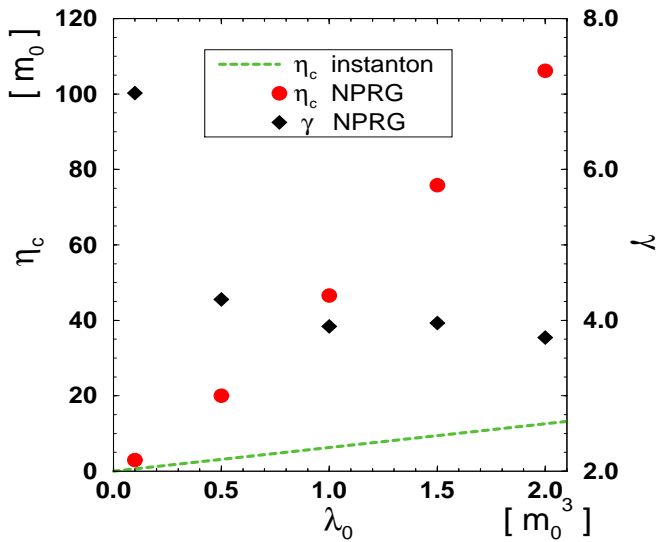}
\caption{\label{fig:eg} $\lambda_0$ dependence of
critical dissipation $\eta_c$ and critical exponent $\gamma$.}
\end{figure}

We find that for every value of $\lambda_0$ $m_{\rm eff}$ decreases as 
$\eta$ becomes large (Fig. \ref{fig:ohm}).
This $\eta$ and $\lambda_0$ dependence of $m_{\rm eff}$ is 
readily understandable by using 
the property of the cutoff interpretation of the Ohmic dissipation.
Remember that the quantum fluctuation of $q$ is suppressed below
the effective infrared cutoff $E_c=\eta$. 
Therefore a larger $\eta$ causes 
a stronger cutoff effect and results in a smaller $m_{\rm eff}$.
For larger $\lambda_0$, the system mass scale is larger, and therefore 
a larger $\eta$ is required 
to effectively cut off the quantum fluctuation.
These behaviors are qualitatively consistent
with those of $\Delta E$ obtained by 
the instanton approximation.
\par
Then, what happens for larger $\eta$?
The expected phenomenon is the quantum-classical transition
characterized by complete disappearance of the tunneling.
The NPRG method may not work precisely in the 
$m_{\rm eff}\to 0$ region because Eq. (\ref{(26)})
becomes singular there and the numerical error increases.
Therefore according to the standard technique of 
analyzing critical phenomena,
we evaluate the critical dissipation $\eta_c$ 
from the diverging behavior of the 
localization susceptibility.
We fit $\chi(\eta)$ with the critical exponent form
 \begin{eqnarray}
\chi=C |\eta-\eta_c|^{-\gamma}.
 \end{eqnarray}
We show an example of fit for $\lambda_0=1.0$ in Fig. \ref{fig:sus}.
We conclude that the localization susceptibility shows divergent 
behavior with a power scaling, which indicates a second order phase 
transition of localization,
often called the quantum-classical transition. 
\par
We obtain the critical dissipation $\eta_c$ and 
the critical exponent $\gamma$ for these $\lambda_0$, which 
are plotted in Fig. \ref{fig:eg}. 
The previous analysis using the dilute gas instanton gives a simple 
relation $\eta_c=2\pi\lambda_0$ 
which is also shown in Fig. \ref{fig:eg} for comparison. 
The NPRG results for $\eta_c$ are systematically large compared to 
those of the instanton.
As for the critical exponent $\gamma$, we observe the universality 
property except for the smallest $\lambda_0=0.1$, where NPRG results 
may not be reliable.
\par
\begin{figure}
\includegraphics[width=80mm,height=80mm]{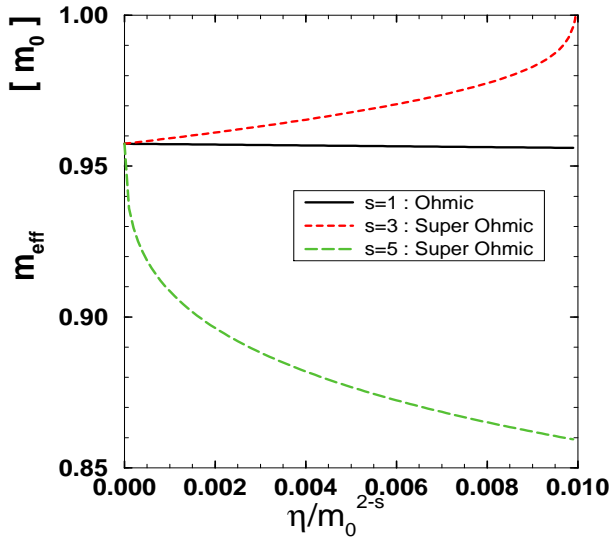}
\caption{\label{fig:dis1} Dissipation dependence of $m_{\rm eff}$ 
for $\lambda_0=0.4[m_0^3]$ with initial cutoff 
$\Lambda_0=100[m_0]$ in the small $\eta$ region. 
We divide $\eta$ by $m_0^{2-s}$ 
in order to make it dimensionless.
}
\end{figure}
\begin{figure}
\includegraphics[width=80mm,height=80mm]{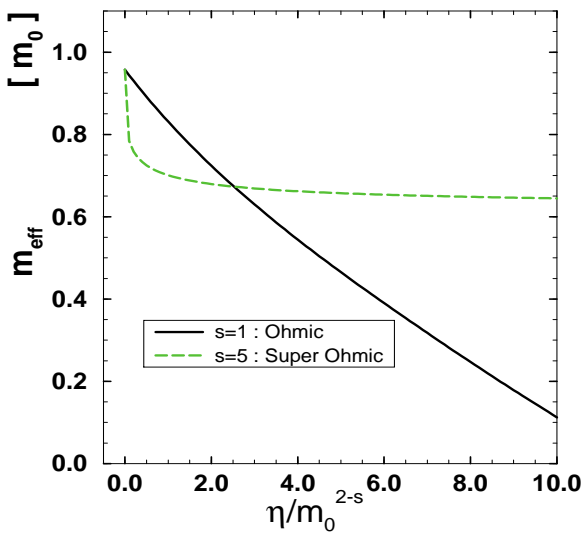}
\caption{\label{fig:dis2} Dissipation dependence of $m_{\rm eff}$ 
for $\lambda_0=0.4[m_0^3]$ with initial cutoff 
$\Lambda_0=100[m_0]$ in the large $\eta$ region.}
\end{figure}

Next we see the difference among various dissipation effects 
at $s=1,3,5$.
The results for $\lambda _0=0.4$ are shown 
in Fig. \ref{fig:dis1} (for small $\eta$) 
and Fig. \ref{fig:dis2} (for large $\eta$).
For small $\eta$, suppression ($s=1,5$) and enhancement (s=3) 
of quantum fluctuations are seen as we expect.  
Note that for the $s=3$ case, since 
the hard ultraviolet cutoff $E_c={1}/{\eta}$ 
exists, the theory is an effective theory defined only below the 
energy scale $E_c={1}/{\eta}$.
Here we use the bare cutoff $\Lambda_0=100$, and  
the $s=3$ result can be obtained only in the region $0\le\eta\le0.01$.
If the system is analyzed perturbatively ($\eta\to 0$), 
such a hard cutoff problem is not recognized \cite{fiss1,fiss2}.
\par
Let us compare the super-Ohmic ($s=5$) case with the Ohmic case.
Both cases suppress the quantum fluctuation and 
the relative strength of the effects depends on $\eta$, 
which turns over at $\eta\sim3.0$ as seen in Fig. \ref{fig:dis2}.
This turnover is explained by comparing the effective cutoff scale 
due to dissipation $E_c=\eta^{-1/(s-2)}$ with the original mass 
scale of the system determined by $m_0$ and $\lambda_0$.

How about the possibility of the quantum-classical transition 
with super-Ohmic dissipation ($s=5,9,. ..$)?
Contrary to the Ohmic case, the dissipation effect corresponds to
the effective ultraviolet cutoff with scale 
$E_c={\eta}^{-1/(s-2)}$.
Taking account of the basic notion that the infrared energy region 
is most significant for the localization transition, 
we may conclude that  
it is impossible for super-Ohmic dissipation
to suppress quantum tunneling completely.
This was expected from naive power counting,
that is, the super-Ohmic dissipation is irrelevant 
in the infrared region.

\section{Summary and Discussion}

We analyzed the Caldeira-Leggett model by 
the nonperturbative renormalization-group method.  
We derived the NPRG equation with a dissipation term 
characterized by $J(\omega)=\eta\omega^s$
within the local potential approximation.
This method enables us to regard dissipation effects 
as various cutoff effects that enhance or suppress the quantum 
fluctuations of the system.
\par
We applied the NPRG equation to analysis of macroscopic quantum
coherence and investigated the quantum-classical transition
due to dissipation. 
For Ohmic dissipation, we observed that   
the localization susceptibility $\chi$ diverges
toward $\eta=\eta_c(\lambda_0)$ with critical scaling, 
and we concluded that $\eta_c(\lambda_0)$ is the critical dissipation 
for the quantum-classical transition.
There must remain, however, many fundamental and subtle problems 
to be studied about the notion of the quantum-classical transition
and critical dissipation.
\par
In this article we did not consider the origin of the environmental
degrees of freedom. 
It is very interesting to regard some higher dimensional fields
as an environment coupled to quantum mechanical particles 
and to study how to treat such a complex system  
by the NPRG method.
Furthermore, it not clear how to discriminate or construct
the target macroscopic collective coordinate 
among the microscopic degrees of freedom.
The NPRG method best fits the purpose of identifying 
the macroscopic effective theory for slow motion
by averaging the quantum fluctuations. 

\begin{acknowledgments}
We would like to thank I. Sawada for fruitful and encouraging 
discussions and suggestions. 
K.-I.A. is partially supported by a Grant-in Aid for
Scientific Research (No.12874029) from the
Ministry of Education, Science and Culture.
\end{acknowledgments}


\begin{thebibliography}{99}
\bibitem{comment2}
In this paper we use the term ``mass'' in the sense of 
quantum field theory, that is, the mass of the exciton.
\end{thebibliography}
\end{document}